\documentclass[conference]{IEEEtran}
\IEEEoverridecommandlockouts
\usepackage{cite}
\usepackage{amsmath,amssymb,amsfonts}
\usepackage{algorithmic}
\usepackage{graphicx}
\usepackage[tight,footnotesize]{subfigure}
\usepackage{textcomp}
\usepackage{xcolor}
\usepackage{optidef}
\usepackage{lipsum}
\usepackage[belowskip=-7pt,aboveskip=-1pt]{caption}
\def\BibTeX{{\rm B\kern-.05em{\sc i\kern-.025em b}\kern-.08em
    T\kern-.1667em\lower.7ex\hbox{E}\kern-.125emX}}

\begin{document}

\title{Robust VAR Capability Curve of DER with Uncertain Renewable Generation\\
}

\author{\IEEEauthorblockN{Aditya Shankar Kar\textsuperscript{1},  Kiran Kumar Challa\textsuperscript{1}, Alok Kumar Bharati\textsuperscript{2}, Ankit Singhal\textsuperscript{3}, Venkataramana Ajjarapu\textsuperscript{1} }
\IEEEauthorblockA{\textit{\textsuperscript{1}Department of Electrical and Computer Engineering, Iowa State University, IA, USA}} 
\IEEEauthorblockA{\textit{\textsuperscript{2}Pacific Northwest National Laboratory,Richland, WA, USA}}
\IEEEauthorblockA{\textit{\textsuperscript{3} Department of Electrical Engineering, IIT Delhi, India}} 

adityask@iastate.edu, kiranc@iastate.edu, ak.bharati@pnnl.gov, sankit@ee.iitd.ac.in, vajjarap@iastate.edu}

\maketitle

\begin{abstract}
Active distribution system with high penetration of inverter based distributed energy resources (DER) can be utilized for VAR-related ancillary services. To utilize the DER flexibility, transmission system operator (TSO) must be presented with the aggregated DER flexibility of distribution system. However, the uncertainty in renewable generation questions the credibility of aggregated capability curve in practice. In this paper, we incorporate the uncertainty into aggregation process to develop a robust capability curve while preserving the real physics (unbalance and lossy nature) of distribution system. 
\textcolor{black}{Statistical inference method is employed to quantify uncertainty in solar generation and quantified uncertainty is integrated into a chance constrained optimal power flow (OPF). It provides the grid operator with the dispatchable aggregated reactive power capability.} The resulting capability curve with the associated probability can be harnessed by the TSO for decision making for both planning and operation.
\end{abstract}


\section{Introduction}
 With an objective of achieving net zero emission by 2050, clean energy generation has gone through vital shift in favour of renewable energy resources\cite{NREL_gen_mix2022}. Continuing the trend, penetration of renewable resources are projected to increase by 150\% to serve 60\% of \textcolor{black}{midcontinent independent system operator} (MISO) load in the coming decade\cite{miso_gen_mix}. Owing to the flexibility in size and process of installation, solar generation has taken significant portion of the projected clean energy production. Hence, the need for conventional synchronous machine based generation are reducing and existing conventional plants are getting retired \cite{miso_gen_mix}. Consequently, there is a decrease in the reactive power reserve, which is a reliable measure of the voltage stability margin \cite{Park2021, Leonardi2008}.

  DER penetrated distribution system (D-system) is no longer a passive network as earlier and is capable of providing ancillary services to grid \cite{Rousis2021,Kontis2021,alok2023}. DERs can offer an alternative solution to the shortage of regional reactive power (VAR) availability. These geographically distributed DERs can be used in Volt/VAR control mode to provide both inductive and capacitive VAR \cite{Zhu2016,singhal2019}. \textcolor{black}{Deriving reactive power support from preinstalled DERs will be significantly cheaper compared to static VAR compensator. Also, the sheer scale of DERs in the D-system can enhance regional VAR availability and affect the voltage stability of electric grid positively}. The aggregated capability of the DERs need to be communicated ahead of time to the TSO by distribution system operator (DSO)/aggregator to be effectively utilized for voltage stability and flexibility.

 In literature, the active power of DERs is aggregated for participating in wholesale energy and service energy markets \cite{Somma2019}. \textcolor{black}{However, our attention is directed at utilizing the DERs' VAR capabilities for enhancing performance of the power grid.}
  
 In this regard, a detailed frame work for VAR provisioning is given in \cite{Singhal2023}. Here, a day ahead aggregated VAR flexibility region (F-R) is derived to support the grid during disturbances \cite{Singhal2023}.
 
 Similarly in \cite{Chen2021}, authors have discussed about robust optimization based method to generate capability curve of active D-system without considering uncertainty in renewable generation. Active and reactive power capabilities of an active D-system are derived in \cite{Silva2018,Zhang2023} assuming deterministic load and renewable generation. The capability curves derived in \cite{Singhal2023,Silva2018,Zhang2023} are contentious as they overlook the inherent uncertainty associated with renewable sources.

 \begin{figure}[b]
    \noindent\fbox{%
        \parbox{\dimexpr\linewidth-2\fboxsep-2\fboxrule\relax}{%
            \begin{tabular}{l}
                This paper is accepted for publication in IEEE PESGM\\  2024. The complete copyright version will be available\\ in IEEE Xplore with published conference proceedings. \\
            \end{tabular}
        }%
    }
\end{figure}
 In recent literature \cite{CCOPF_Zhang,Tan2020,potential_reactive_power}, few researchers have attempted to derive the capability of D-system including the uncertainty in renewable generation. In \cite{CCOPF_Zhang}, uncertainty is included while deriving an aggregated active power capability of DER for real time market applications and does not provide any information on the VAR capability. \textcolor{black}{In \cite{Tan2020}, uncertainty is considered in the maximum active power generation ($P_{TVPP}^{g max}$) of a technical virtual power plant (TVPP). While deriving an aggregated active and reactive power capability region, the active power is assumed to be flexible in the range of 0 to $P_{TVPP}^{g max}$. Since flexibility in reactive power \cite{Tan2020} results in active power curtailment and consumer revenue, it is not in the best interests of consumers.} \textcolor{black}{A probable set of optimal aggregated reactive power curves are presented in \cite{potential_reactive_power}. The proposed method cannot be directly used by the TSO for decision making. The approach mentioned in \cite{potential_reactive_power} is computationally cumbersome with requirement of very high number of random samples. There are examples of solving probabilistic optimization models for accounting for renewable uncertainty \cite{Zhang_Trans_2019,Stankovic2020}, however, it is not feasible to employ these methods in practice due to various limitations.}

\textcolor{black}{To address the concerns mentioned in the previous paragraph, there is a need for robust VAR capability curve of a D-system. This curve should furnish a quantifiable numerical result of aggregated reactive power with associated confidence factor. In this paper, we define a robust VAR flexibility region (F-R) of D-system as $[\overline{Q}^{sub}_{deamnd}, \underline{Q}^{sub}_{deamnd}]$ at the substation (``$\overline{Q}^{sub}_{deamnd}$" and ``$\underline{Q}^{sub}_{deamnd}$" represent maximum inductive and capacitive VAR support from D-system respectively).}

DERs in D-system can be controlled to achieve aggregated reactive power demand in the F-R $[\overline{Q}^{sub}_{deamnd}, \underline{Q}^{sub}_{deamnd}]$. 
\textcolor{black}{The novelty of the proposed method is summarized below:
\begin{itemize}
\item   Uncertainty in solar power generation data\cite{NREL_Data} is quantified using statistical tools in a mathematically tractable way which can be incorporated into aggregation process.
\item	A credible DER VAR capability estimate is proposed which will quantify amount of aggregated reactive power ahead of time with given probability/risk factor so that the transmission operator can incorporate VAR capability into their decision-making algorithms. The proposed method ensures that the day ahead calculated aggregated reactive power will be dispatchable for service during the hour of operation.
\item \textcolor{black}{We utilized domain knowledge to reformulate the probabilistic optimization model via gaussian reformulation which is easier to solve.}
\item  To establish a practically valid capability curve, we employ a 3-phase unbalanced D-system model,  as unbalanced distribution of DERs can significantly affects the VAR capability. The determination of inverter size is guided by adherence to the IEEE 1547-2018 requirements.
\end{itemize}}

\section{Quantification of Uncertain Solar Generation}\label{section:Uncertainty in solar}

The uncertainty in solar irradiance affects the generation of active power. Hence, it affects the amount of reactive power derived out of the DER. But from the grid operator prospective, uncertainty in reactive power generation needs to be quantified for operation and decision making.

 In \cite{NREL_Data}, hourly forecast data and actual data of solar generation for a year are presented for integration and operation study. We used historical data to quantify the error present in the forecast data. The objective is to create a model which will provide an expected value of the error for a given solar generation forecast, but not to upgrade the forecasting process. Finally for aggregation, it should be appropriate to use the forecast data in conjunction with the quantified uncertainty in terms of error in solar generation.

To create a model, errors in 8760 solar generation data points are required to be appropriately consolidated or grouped to provide information on errors for corresponding solar generation prediction value. Consolidation can be done in many ways. one of the consolidation methods is to group all the data that correspond to the same time of day. The primary drawback of this method is seasonal variation of the sunlight availability w.r.t time of day. Thus, it produces large variance. Instead, we observe the error in similar solar forecast value in the historical data. The confidence interval of error is much tighter in the later method compared to the previous one. Here, relative error is larger for lower generation forecast and it is smaller comparatively in case of large generation forecast. It signifies that absolute error magnitude remains similar to the prediction range. This can be considered as an indicator of accuracy of the prediction method and help operators to incorporate the quantified uncertainty in decision making.

\begin{figure}

 \hfill
\subfigure[]{\includegraphics[width=4.3cm]{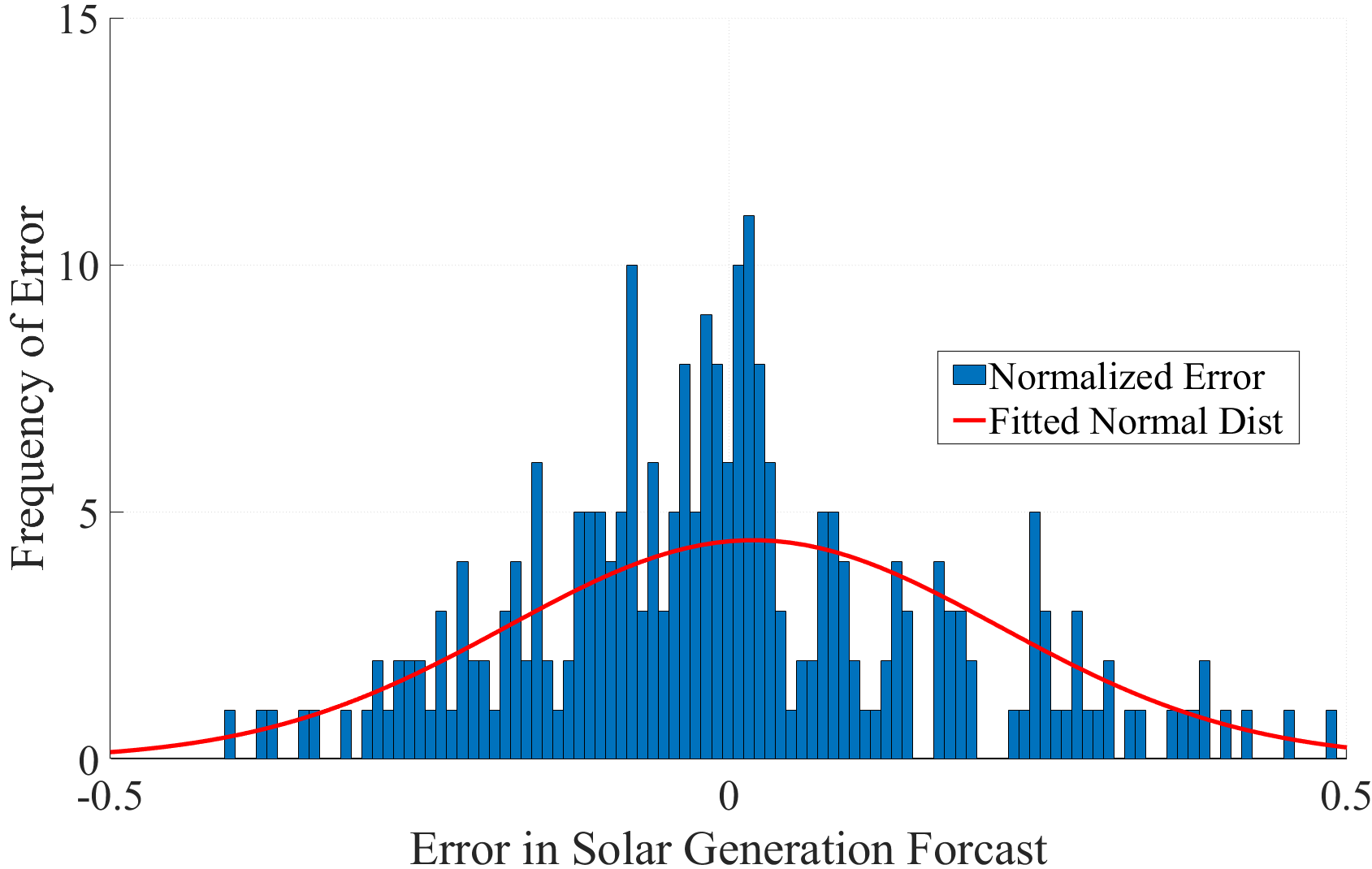}\label{fig:N_dist_of_error1}}
\hfill
\subfigure[]{\includegraphics[width=4.3cm]{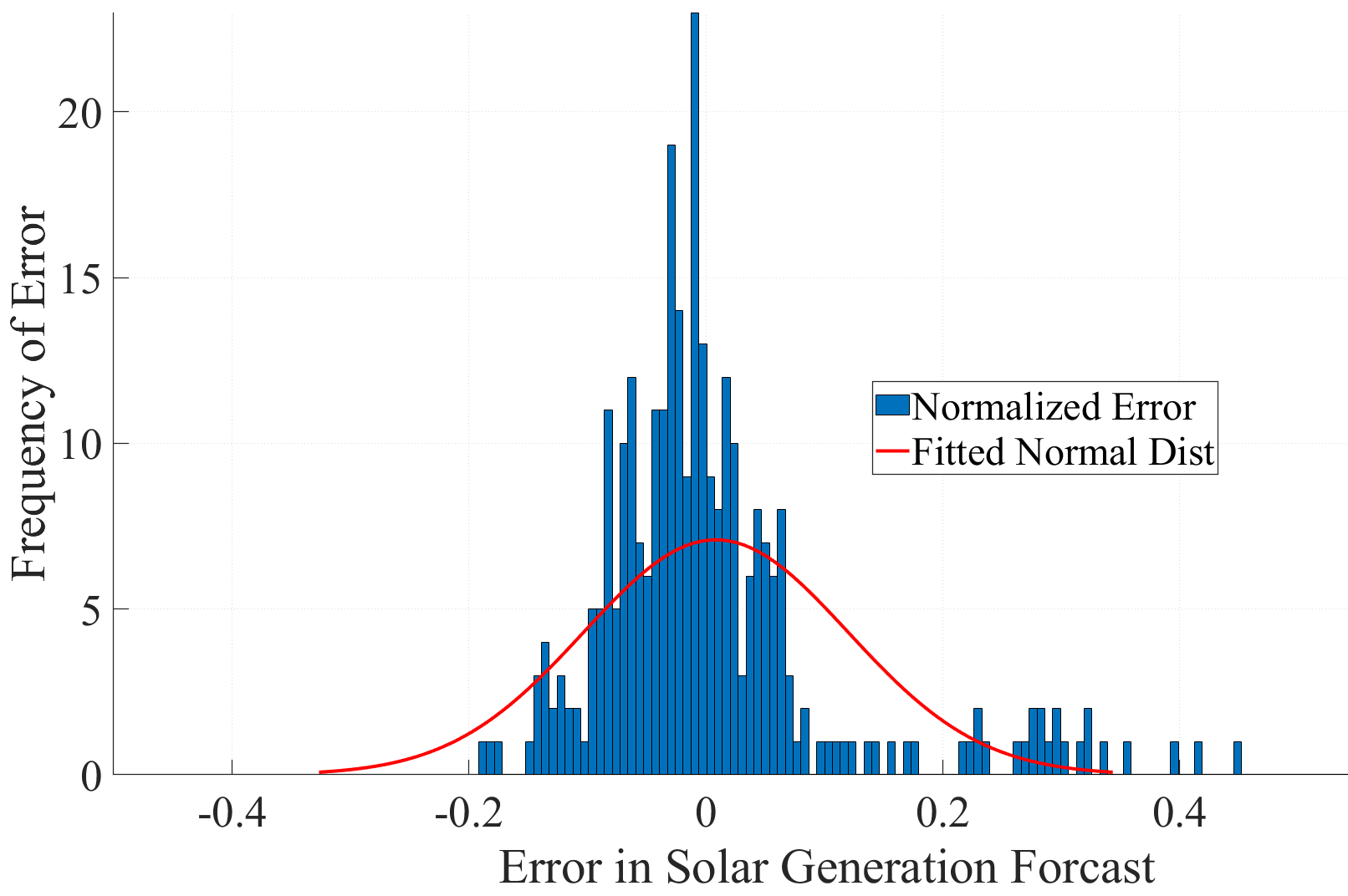}\label{fig:N_dist_of_error2}}
\hfill
\caption{Histogram and fitted normal distribution of the calculated error}
\label{fig:Histogram_dist_of_error}
\end{figure}

Using the data in  \cite{NREL_Data}, error in forecast is calculated and the histogram can be plotted to observe its approximate distribution. In Fig.\ref{fig:N_dist_of_error1} \& \ref{fig:N_dist_of_error2}, histogram of relative error is plotted for predicted generation percentage of 50\% and 62.5\% respectively. We observe histograms of errors to be close to both normal distribution as well as student's t-distribution. As in most of the cases degree of freedom (number of observation) of error is more than 20, we fit the error to a normal distribution for the forecast values as shown in Fig.\ref{fig:Histogram_dist_of_error}. Details of relative error calculation (normalized error) with normal distribution fitting are articulated below.

\begin{itemize}
\item For accuracy, the night time data are removed where the forecasted value is zero.
\item Normalize solar forecast value w.r.t the installed capacity.
\item Calculate the absolute error and relative error w.r.t forecast value for all the data points.
\item Group the similar normalized forecast value and their error together.
\item Plot the histogram of relative error to see the distribution.
\item Fit a normal distribution to the relative error and calculate the mean and standard deviation of the distribution. 
\end{itemize}

\vspace{0.3cm}
In Table \ref{table:error_quant_table}, the mean($\mu$) and standard deviation ($\sigma$) of normal distribution is given for the relative error corresponding to predicted generation percentage of installed capacity for few cases.

\begin{table}[!h]
    \centering
    \caption{ Quantified error in solar generation}
    \begin{tabular}{|l|p{15mm}|p{15mm}|p{15mm}|}
    \hline
        ~ & Mean \newline Error & Standard \newline Deviation & Prediction \newline Percentage \\ \hline
        1 &  0.2341 & 0.6543 & 0.0464 \\ \hline
        2 & -0.0169 & 0.5751 & 0.1234  \\ \hline
        3 & -0.0863 & 0.3702 & 0.2066  \\ \hline
        4 &  0.0044 & 0.4007 & 0.2936   \\ \hline
        5 & -0.0529 & 0.2629 & 0.3762 \\ \hline
        6 & -0.0299 & 0.2525 & 0.4598 \\ \hline
        7 & -0.0039 & 0.1598 & 0.5422  \\ \hline
        8 &  0.0083 & 0.1116 & 0.6260  \\ \hline
    \end{tabular}
    \label{table:error_quant_table}
\end{table}

\section{System Modelling}
Accuracy and validity of the capability curve for D-system depends entirely on its mathematical description. In the following section, model of inverter based DERs and linearized network are discussed with the pursuit of including them in the OPF formulation.

\subsection{Description of Hardware and their Limits }\label{section: hardware Description}
\begin{figure}
 \centering

\includegraphics[width=9cm,height=12cm,keepaspectratio]{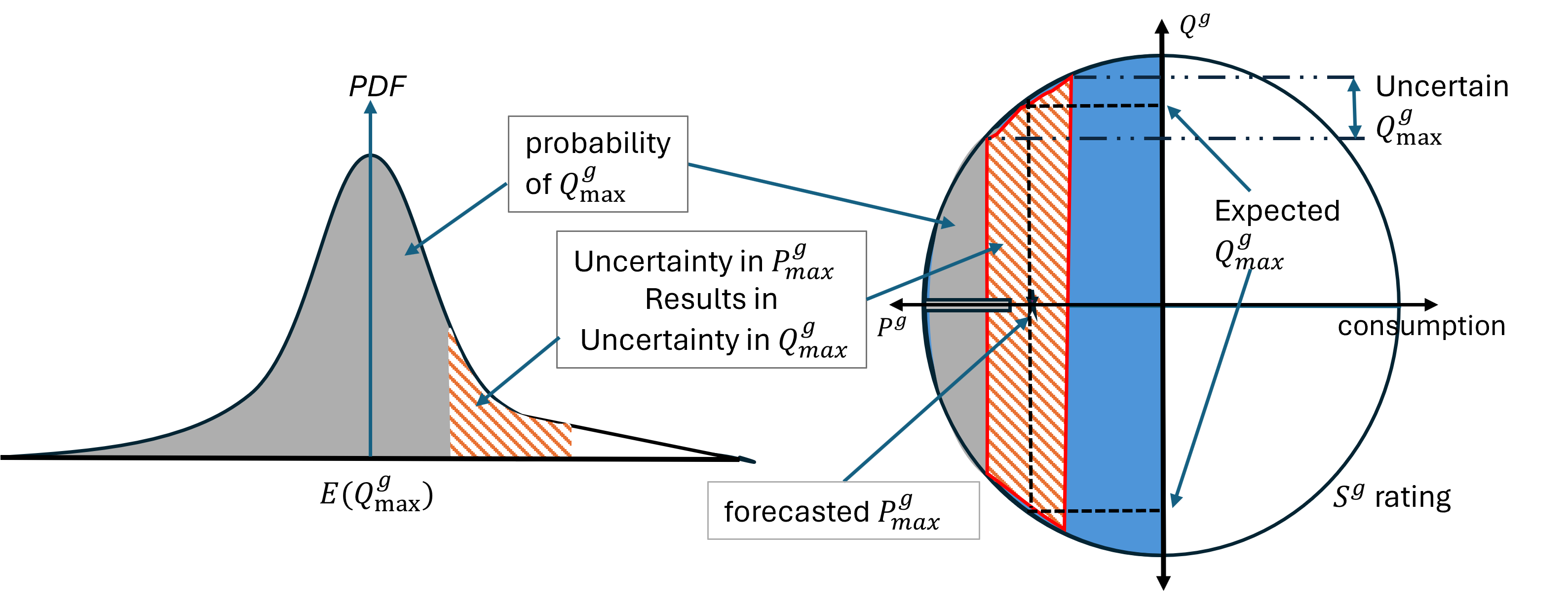}
\caption{ \textcolor{black}{Capability curve of inverter-based DER and Associated Uncertainty}}
\label{fig:hardware figure}
\setlength{\belowcaptionskip}{-1cm}
 \end{figure}

The P-Q capability of a single inverter can be plotted on a 2-D plane. If the x-axis represents the active power output and y-axis represents the reactive power output of the inverter, the capability curve can be represented as circle on P-Q plane as shown in Fig.\ref{fig:hardware figure}. Any point inside the boundary of the circle can be an operation point with appropriate control. However, the operating point is decided by the maximum active power output of DER. Depending on the active power output, the reactive power is limited due to the hardware limit of inverter of DER. As there is cost associated with the active power consumption, curtailing the active power may not be of interest for the owners of DERs. 

But the recent IEEE 1547-2018 standard, mandates that DER must not deliver more than 90\% of its KVA rating during operation. So, the DER can provide reactive power up to 44\% of the inverter KVA rating to support the grid operation. Moreover, if the active power generation is lesser than more reactive power can be generated out of the IBRs as presented in \textcolor{black}{Fig.\ref{fig:hardware figure}}. Finally, operating point constraint is given in (\ref{eq: uncertainty eq1}). 
\vspace{-0.15cm}
\begin{equation}
{{q_{j}^{g}}^2 \leq {S_{j}^{g}}^2 -\Tilde{{p_{j}^{g}}}^2} \label{eq: uncertainty eq1}
\end{equation}
where $\Tilde{p_{j}^{g}} $ represents uncertain solar generation $p_{j}^{g}$ at node \textit{j}.

\subsection{Linear Model }\label{linear_model}
A linear model of D-system is used in OPF formulation which is an extension of \textit{LinDistFlow} to unbalance 3 phase circuit \cite{Singhal2023,Arnold2016}. The goal is to determine a linear relationship between the squared voltage term and power injection at D-system nodes to use in linear optimization. Applying KVL \& KCL to the adjacent node of the 3phase unbalance network, we can write  (\ref{lin_volt_eq1}), where $P_{j} = [P_{a}  P_{b}  P_{c}]_{j}^T$ \& $Q_{j} = [Q_{a}  Q_{b}  Q_{c}]_{j}^T$ represent the active and reactive power entering at node \textit{j} for phase \textit{a,b,c} and $V_{j}$ represents the 3phase voltage vector for node \textit{j}. $Z_{ij}^p$ and $Z_{ij}^q$ are three phase impedance coefficient matrix of the line connecting node \textit{i\&j}. 
\vspace{-0.15cm}
\begin{equation}
V_{i}V_{i}^* = V_{j}V_{j}^*-Z_{ij}^pP_{j}-Z_{ij}^qQ_{j} \label{lin_volt_eq1}
\end{equation}

  Using the connectivity matrix '$M$', we can represent (\ref{lin_volt_eq1}) in compact form given in (\ref{lin_volt_eq2}). The squared voltage term is written with the variable $Y_{i} = V_{i}V_{i}^*$.
\begin{equation}
\begin{bmatrix}  M_{0} & M^{T}  \end{bmatrix} \begin{bmatrix}  Y_{0} & Y  \end{bmatrix}^{T}=-Z_{ij}^pP_{j}-Z_{ij}^qQ_{j} \label{lin_volt_eq2}
\end{equation}
\vspace{0.1cm}
However, the voltage dependencies of the loads increase the complexity. Considering  $a_{\phi,j}^{0}$ \& $a_{\phi,j}^{1}$ are the fraction of load which are constant power and voltage dependent respectively at phase $\phi$ and  node \textit{j}. The load can be represented as 
\begin{equation}
p_{\phi,j}^{l}(V_{\phi,j}) = p_{\phi,j}^{l}(a_{\phi,j}^{0}+ a_{\phi,j}^{1} V_{\phi, j}^{2}) \label{load eq1}
\end{equation}
Also, the difference in power entering at a node and leaving for connected adjacent node, can be equated  to net load demand at each node. Net load is difference in load and generation at that node. This power balance is represented in (\ref{Load Gen eq2}).
\begin{equation}
\begin{aligned}
MP= p^{g} -  (p^{l}a_{\phi,j}^{0}+ diag(p^{l})Ya_{\phi,j}^{1} ) \\
MQ= q^{g} -  (q^{l}a_{\phi,j}^{0}+ diag(q^{l})Ya_{\phi,j}^{1} )
\end{aligned}
 \label{Load Gen eq2}
\end{equation}
Using \eqref{lin_volt_eq2} \& \eqref{Load Gen eq2} the expression of squared voltage is given in \eqref{lin_volt_eq3}.
\begin{equation}
Y = K^{-1} [R^{eq} (p^{g}-p^{l}a_{\phi,j}^{0}) + X^{eq}(q^{g}-q^{l}a_{\phi,j}^{0}) - M^{-T}M_{o}Y_{O}] \label{lin_volt_eq3}
\end{equation}
 Where 
\begin{equation}
\begin{aligned}
R^{eq} = -M^{-T} Z_{D}^{P}M^{-1}  \\
X^{eq} = -M^{-T} Z_{D}^{q}M^{-1} \\
K = I_{3N} + R^{eq}diag(p^{l})a^{1}+  X^{eq}diag(q^{l})a^{1}\\
\end{aligned}
\label{lin_volt_eq4}
\end{equation}

\subsection{Solar Generation with Associated Probability}\label{Erroneous_Generation}

Uncertainty in solar generation is quantified by associating the probability to the predicted values of solar power generation. The procedure followed in embedding this information into the reactive power capability curve of DER is explained in this section to provide more accurate and actionable information for TSO. Let us consider the DER capability as discussed in Section \ref{section: hardware Description}. Taking the expected value on both sides of (\ref{eq: uncertainty eq1}) and representing the entire formulation by an equivalent deterministic model, as in (\ref{eq: uncertainty eq2}), is a convenient way to deal with uncertainty. 
\begin{equation}
E({q_{j}^{g}}^2) \leq E({S_{j}^{g}}^2 -\Tilde{{p_{j}^{g}}}^2) \newline
= {S_{j}^{g}}^2 - E(\Tilde{{p_{j}^{g}}}^2) \label{eq: uncertainty eq2}
\end{equation}

\textcolor{black}{From  Fig.\ref{fig:hardware figure} and equation (\ref{eq: uncertainty eq1}) \& (\ref{eq: uncertainty eq2}), it can be observed that,} if the actual $p^g_j$ is higher than $E(\Tilde{{p_{j}^{g}}})$ with positive error, then the room for maximum possible reactive power generation ($q_{j}^{g}$) at node \textit{j} will be lesser than calculated value. On other hand, if the actual $p^g_j$ goes below the $E(\Tilde{{p_{j}^{g}}})$, then the maximum possible $q_{j}^{g}$ will be higher than calculated value. Hence the aggregated reactive power of D-system holds good, if $p^g_j<E(\Tilde{{p_{j}^{g}}})$ but it does not remain valid for $p^g_j>E(\Tilde{{p_{j}^{g}}})$. So with the expected value of $E(\Tilde{{p_{j}^{g}}})$, inequality in (\ref{eq: uncertainty eq2}) holds good with probability of 0.5.

Validity of capability curve can be improved under uncertain scenario by including the chance constrained formulation given in (\ref{eq: uncertainty eq3}). Essentially  (\ref{eq: uncertainty eq3}) ensures that hardware constrained (\ref{eq: uncertainty eq1}) holds with a probability $1-\alpha$ where $\alpha$ is the risk factor.

\begin{equation}
\mathbb{P}({q_{j}^{g}}^2 \leq {S_{j}^{g}}^2 -\Tilde{{p_{j}^{g}}}^2) \geq 1-\alpha
 \label{eq: uncertainty eq3}
\end{equation}

In practice, it is preferred to satisfy (\ref{eq: uncertainty eq3}) for very high probability with lower ($\alpha$) risk factor. Considering $\widehat{p_{j}^g} = E(\Tilde{{p_{j}^{g}}}) + \mu_{error} + \sigma_{error} $ and $\widehat{p_{j}^g} = E(\Tilde{{p_{j}^{g}}}) + \mu_{error} + 2\sigma_{error} $ respectively, we can ensure ( \ref{eq: uncertainty eq3})  with probability $ \mathbb{P} = 0.84 $  and $  \mathbb{P} = 0.976$. For ensuring the validity of ( \ref{eq: uncertainty eq3}) with a probability of $\mathbb{P}$, the value of active power $p_{j}^g$ should be considered according to (\ref{eq: uncertainty eq4}).

\begin{equation}
\begin{aligned}
\widehat{p_{j}^g} = E(\Tilde{{p_{j}^{g}}})  + \mu_{error} + z\sigma_{error} 
\label{eq: uncertainty eq4}
 \end{aligned}
\end{equation}
where $z = \Phi^{-1}(\mathbb{P})$ and $\Phi(z) = \int_{-\inf}^{z}  \frac{1}{\sqrt{2\pi}} e^{{\frac{-x^{2}}{2}}} \,dx$ represents the cumulative distribution function of the standard normal.

\section{OPF Formulation}
VAR flexibility region (F-R) of a D-system can be defined as the maximum reactive power that can be absorbed (lagging VAR) or generated (leading VAR) by DERs. To calculate the available VAR F-R at the substation, we need to minimize and maximize the net VAR drawn from the substation. For a given solar generation $p_{j}^{g}$ from inverter, we can use rest of VA capacity to generate the $q_{j}^{g}$. Adding the  $q_{j}^{g}$ at each node algebraically, can potentially violate the D-system constraint. Hence, an OPF is formulated to minimize and maximize the net VAR injection by D-system as given in (\ref{eqn: optimization_obj}). The objective function (\ref{eqn: optimization_obj}) is the difference between the VAR generation and demand at the D-system node. In this OPF formulation, the solar power generation, active and reactive power demand of load, are given as input and the reactive power generation $q_{j}^{g}$ of individual DER is the output of the optimization module.
\begin{mini!}|s|
{q_{j}^{g}}{\sum_{i=1}^{N}{q_{j}^{g}}-\sum_{i=1}^{N}{q_{j}^{l}}}
{}{}
\label{eqn: optimization_obj}
\addConstraint {Y = K^{-1}[R^{eq}(\Tilde{p^{g}}-p^{l}a^{0})+ X^{eq}(q^{g}-q^{l}a^{0})+ 1 v_{0}^{2} ]}
\label{eqn: optimization_const1}
\addConstraint {\underline{y_{j}} \leq y \leq \overline{y_{j}}}
\label{eqn: optimization_const5}
\addConstraint {p_{j} = \Tilde{p_{j}^{g}}-p_{j}^{l}(a_{j}^{0}+a_{j}^{1}y_{j})}
\label{eqn: optimization_const2}
\addConstraint {q_{j} = {q_{j}^{g}}-q_{j}^{l}(a_{j}^{0}+a_{j}^{1}y_{j})}
\label{eqn: optimization_const3}
\addConstraint {\mathbb{P}({{q_{j}^{g}}^2 \leq {S_{j}^{g}}^2 -\Tilde{{p_{j}^{g}}}^2}) \geq 1 - \alpha}
\label{eqn: optimization_const4}
\end{mini!}

In the formulation, constraint (\ref{eqn: optimization_const1}) is a linear function of squared voltage terms which represents 3-phase linearized power flow. The upper and lower bound of the each nodal voltage is set to 1.05 and 0.95 respectively in (\ref{eqn: optimization_const5}). Active and reactive power balance in D-system are presented in (\ref{eqn: optimization_const2}) \& (\ref{eqn: optimization_const3}). The hardware limits are given in (\ref{eqn: optimization_const4}). Here, OPF is formulated with uncertain solar generation. With $p_{j}^{g}$ is represented as a random variable $\tilde{p_{j}^{g}}$, the constraint (\ref{eqn: optimization_const3}) is modeled as chance constrained problem. However, for solving the optimization problem, solar power generation($\widehat{p_{j}^{g}}$) with associated probability ($\mathbb{P}$) is calculated as described in  (\ref{eq: uncertainty eq4}). This $\widehat{P_{j}^{g}}$ can be used to ensure equation (\ref{eqn: optimization_const4}) with probability $\mathbb{P} \geq 1-\alpha $. With all other deterministic parameters, we replace $\tilde{p_{j}^{g}}$ with $\widehat{P_{j}^{g}}$ in (\ref{eqn: optimization_const1}), (\ref{eqn: optimization_const2}) for a Gaussian reformulation of OPF.

\section{Test System Description and Result}
\begin{figure}
\hfill
\subfigure[]{\includegraphics[width=4.3cm]{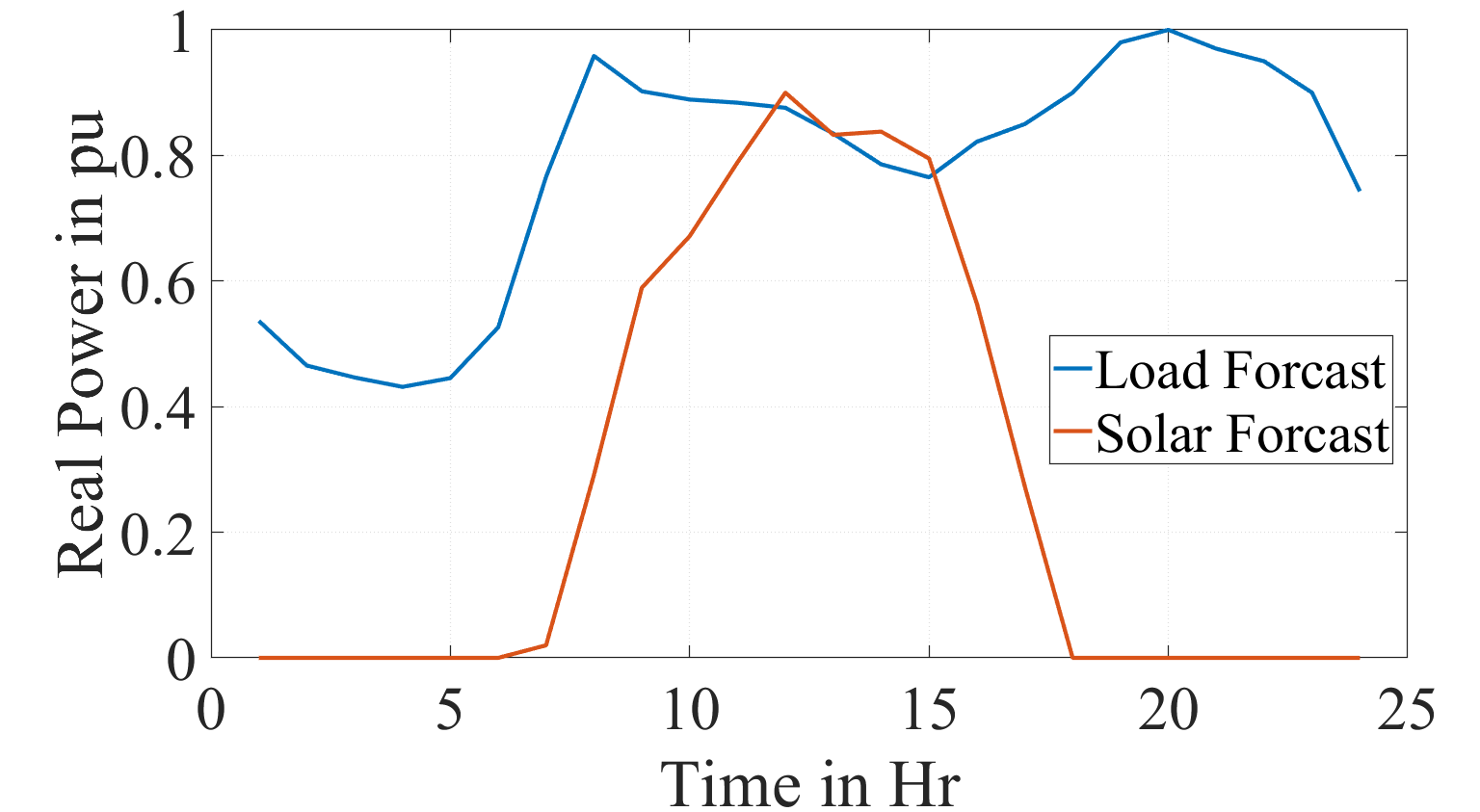}\label{fig:Deterministic_Load}}
\hfill
\subfigure[]{\includegraphics[width=4.3cm]{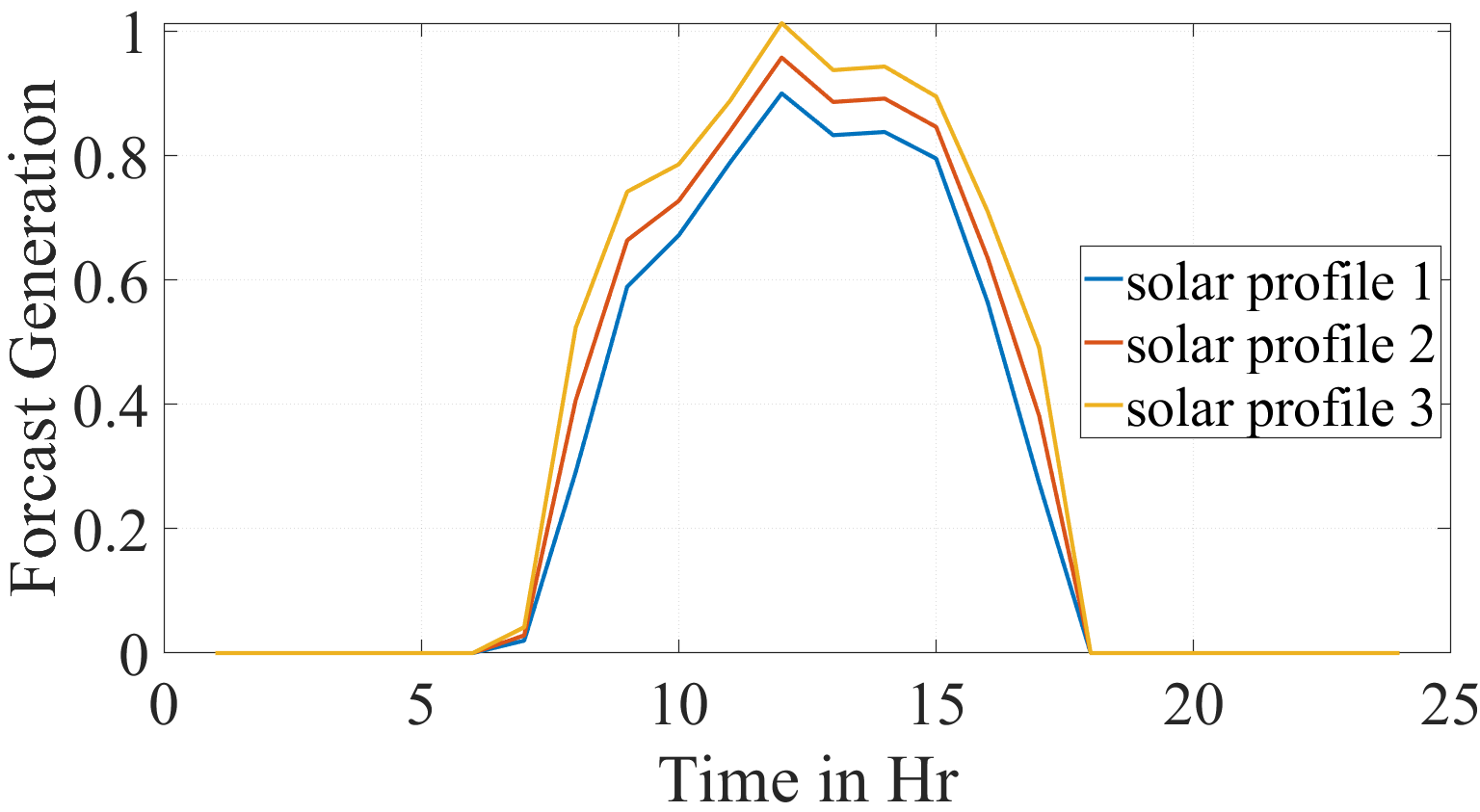}\label{fig:Predicted_curve}}
\hfill
\caption{(a) Day ahead load and solar generation forecast, (b) Generated solar profile with associated probability}
\label{fig:Solar profiles}
\end{figure}

We have considered IEEE 123 node 3-phase unbalanced D-system with 3.5 MW demand. Solar generation penetration is considered as 90\% of the peak load. Equal capacity of DERs are connected at all the nodes in D-system. Inverter ratings are considered to be 10\% higher than the peak solar generation level to be consistent with IEEE-1547. The solar and load profiles for 24 hours are given in Fig.\ref{fig:Deterministic_Load}.

As discussed in section \ref{section:Uncertainty in solar}, we have quantified the error in solar power prediction to be Gaussian distribution representing the error in terms of mean error and standard deviation. The deterministic solar generation profile presented in Fig.\ref{fig:Deterministic_Load}, is considered as the expected solar generation value for 24 hour. Matching it with the prediction percentage of Table \ref{table:error_quant_table}, the mean error and it's associated stand deviation is decided. For desired probability value "$\mathbb{P}$"($\mathbb{P} = 0.84$ and $\mathbb{P} = 0.976$), the synthetic solar power generation generation profile is created using (\ref{eq: uncertainty eq4}) as shown in Fig.\ref{fig:Predicted_curve}. The forecasted solar power generation is presented as solar profile 1. Solar profile 2 and solar profile 3 are calculated with associated probability $\mathbb{P} = 0.84$ and $\mathbb{P} = 0.976$. 

\begin{figure}
\centering
\includegraphics[width=8cm,height=10cm,keepaspectratio]{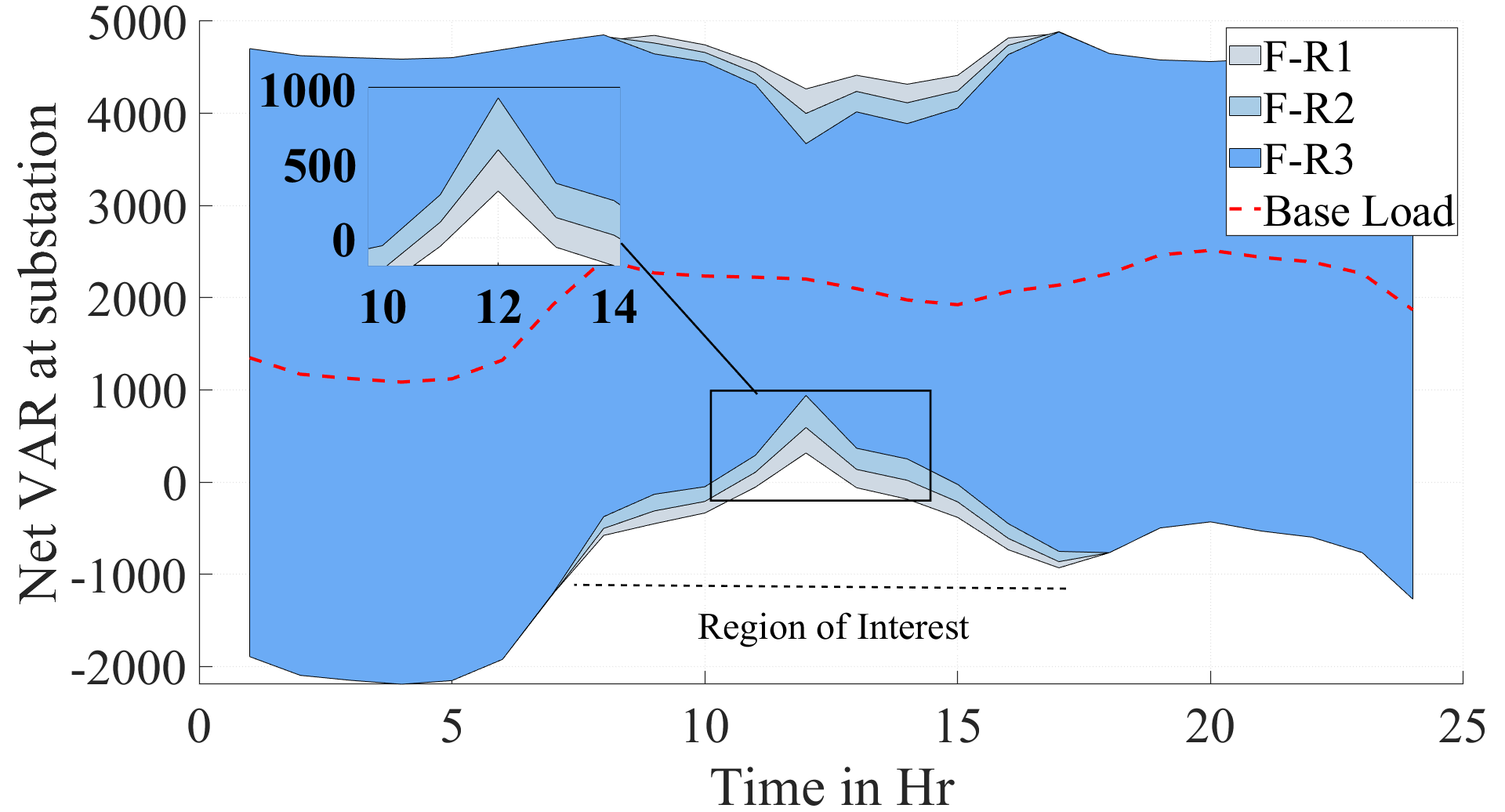}
\caption{VAR flexibility regions (F-R) of D-system  with different probability}
\label{fig:Flexibility_Curve}
\end{figure}

The solar generation profiles presented in Fig.\ref{fig:Predicted_curve} are used for solving the Gaussian reformulated OPF \cite{CCOPF_Zhang}. As the formulated OPF is convex with linear constraints and objective function, it  can be solved efficiently using the standard routines. The resulting aggregated capability curve of D-system is presented in Fig.\ref{fig:Flexibility_Curve}, where the flexibility region 1 (F-R1), F-R2, F-R3 corresponds to the solar profile 1, 2, 3 respectively. Hence, the probability of F-R remaining valid, can be attributed to the probability ($\mathbb{P}$) associated with the synthetic solar profile. 
The aggregated F-R2 and F-R3 are valid with the probability of 0.84 and 0.97 respectively. With the dotted line in Fig.\ref{fig:Flexibility_Curve} as a base load, the D-system can be operated with in the shaded region. During the nighttime, as the solar generation is absent, we are certain about the VAR capability curve of D-system. With the variation in solar generation during the day, VAR capability curve of D-system changes. Hence, day time of capability curve is region of interest as shown in Fig.\ref{fig:Flexibility_Curve}. The aggregated capability curve during the period of interest, is presented in Table \ref{table:Capability_Curve} with the associated probability. The maximum and minimum operating points are indicated with respect to time of the day for the predicted load and solar generation. The -ve VAR in Table \ref{table:Capability_Curve}, can be considered as the reactive power injection from the D-system to the transmission system (T-system). In Table \ref{table:Capability_Curve}, the variation in $\underline{Q}^{sub}_{min}$ at $12^{th}$  hour is 624 (939-315 =624) VAR for a single distribution network. However, as seen by the TSO, this variation due to uncertainty will be multi-fold with 10s to 100s distribution networks downstream at a given T-system. 

\vspace{0.3cm}
\begin{table}[!h]
    \centering
    \caption{ Aggregated capability curve of the D-system}
    \begin{tabular}{|l|l|l|l|l|l|l|}
        \hline
       ~ & \multicolumn{2}{|c|} {Base case \newline P =0.5}  & \multicolumn{2}{c|} {P =0.86}  & \multicolumn{2}{c|} {P =0.97}    \\ \hline
        hr & $\overline{Q}^{sub}_{max}$ & $\underline{Q}^{sub}_{min}$  &  $\overline{Q}^{sub}_{max}$ & $\underline{Q}^{sub}_{min}$ &  $\overline{Q}^{sub}_{max}$ & $\underline{Q}^{sub}_{min}$  \\ \hline
       6 & 4691 & -1923 & 4691 & -1923 & 4691 & -1923  \\ \hline
       7 & 4762 & -1210 & 4768 & -1213 & 4777 & -1204  \\ \hline
       8 & 4788 & -578 & 4834 & -503 & 4851 & -375  \\ \hline
       9 & 4846 & -452 & 4762 & -313 & 4645 & -132  \\ \hline
       10 & 4742 & -335 & 4660 & -210 & 4557 & -50  \\ \hline
       11 & 4546 & -55 & 4438 & 106 & 4311 & 290  \\ \hline
       12 & 4264 & 315 & 3998 & 591 & 3670 & 939  \\ \hline
       13 & 4413 & -61 & 4237 & 137 & 4015 & 367  \\ \hline
       14 & 4315 & -185 & 4114 & 19 & 3888 & 250  \\ \hline
       15 & 4411 & -383 & 4243 & -215 & 4056 & -25  \\ \hline
       16 & 4817 & -734 & 4739 & -611 & 4637 & -451  \\ \hline
       17 & 4865 & -930 & 4886 & -863 & 4882 & -752  \\ \hline
       18 &  4648 & -766 & 4648 & -766 & 4648 & -766  \\ \hline
    \end{tabular}
    \label{table:Capability_Curve}
\end{table}

\vspace{-0.2cm}

\section{Discussion and Conclusion }
In this paper, we have presented a methodology for including uncertainty in the DER aggregation process. The resulting aggregated capability curve is expected to remain valid during uncertain solar generation scenarios. The information from Table \ref{table:Capability_Curve} (communicated a day ahead to TSO) along with current solar insolation measurements will help TSO in making a confident decision to request a certain amount of VAR from DSO. The robust capability curve will be essential as the variation due to uncertainty is large. In absence of uncertainty consideration, the TSO cannot make a confident decision. A robust capability curve will provide opportunity for TSO to enhance the efficiency of the transmission system and congestion management during normal operation. It also enables effective reactive power management during contingencies. As the unbalance in D-system and inverter sizing are considered in aggregation process, the developed aggregated capability curve will be dependable for TSO \& DSO decision making. This additional knowledge of aggregated DER capability enables utilities for peak VAR management. In future work, we plan to include the uncertainty in load in active distribution and operating condition of DERs.

\bibliography{my_library.bib}
\bibliographystyle{IEEEtran}

\end{document}